\documentclass[12pt,a4paper]{iopart}
\pdfoutput=1
\usepackage{iopams,setstack,mathrsfs,amsthm,cite,enumerate,float,graphicx,enumerate,ytableau}
\usepackage[utf8]{inputenc}
\usepackage[T1]{fontenc}
\usepackage{lmodern}
\usepackage[colorlinks,linkcolor=blue,citecolor=blue,urlcolor=blue]{hyperref}
\newcommand{\al}{\alpha}
\newcommand{\be}{\beta}
\newcommand{\de}{\delta}
\newcommand{\ep}{\epsilon}
\newcommand{\vep}{\varepsilon}

\newcommand{\la}{\lambda}
\newcommand{\om}{\omega}
\newcommand{\si}{\sigma}

\newcommand{\vp}{\varphi}

\newcommand{\ze}{\zeta}
\newcommand{\Om}{\Omega}
\newcommand{\bde}{\boldsymbol{\delta}}

\newcommand{\bk}{\mathbf{k}}

\newcommand{\bbs}{\mathbf{s}}

\newcommand{\bT}{\mathbf{T}}

\newcommand{\by}{\mathbf{y}}

\newcommand{\bsi}{{\boldsymbol{\si}}}

\newcommand{\NN}{{\mathbb N}}
\newcommand{\RR}{{\mathbb R}}
\newcommand{\ZZ}{{\mathbb Z}}
\newcommand{\cF}{{\mathcal F}}
\newcommand{\cH}{{\mathcal H}}

\newcommand{\cN}{{\mathcal N}}
\newcommand{\cP}{{\mathcal P}}

\newcommand{\cZ}{{\mathcal Z}}
\newcommand{\ket}[1]{|#1\rangle}

\newcommand{\mss}{\kern 1pt}

\renewcommand{\le}{\leqslant}
\renewcommand{\ge}{\geqslant}
\newcommand{\tends}[1]{\bbuildrel{\hbox to 2em{\rightarrowfill}}_{#1}^{}}

\newcommand{\iu}{\mathrm i}
\newcommand{\diff}{\mathrm{d}}

\newcommand{\su}{\mathrm{su}}

\newcommand{\gl}{\mathrm{gl}}
\newcommand{\en}{\enspace}

\newcommand{\pdf}[2]{\frac{\partial #1}{\partial #2}}

\newcommand{\Int}[1]{\,\mathop{\!#1}\limits^{\lower1ex\hbox{$\scriptstyle\circ$}}{}}

\newcommand{\db}{b^\dagger}
\newcommand{\dc}{c^\dagger}
\newcommand{\da}{a^\dagger}

\newcommand{\df}{f^\dagger}
\newcommand\vru{\vrule width0pt height11pt depth6pt}

\theoremstyle{remark}

\let\tfrac\case
\let\eqref\eref

\newcommand{\binom}[2]{{#1\choose #2}}

\newcommand{\dfrac}{\displaystyle\frac}
\eqnobysec
\newcommand{\bN}{\mathbf N}

\def\clap#1{\hbox to 0pt{\hss#1\hss}}

\def\mathrlap{\mathpalette\mathrlapinternal}
\def\mathclap{\mathpalette\mathclapinternal}

\def\mathrlapinternal#1#2{%
           \rlap{$\mathsurround=0pt#1{#2}$}}
\def\mathclapinternal#1#2{%
           \clap{$\mathsurround=0pt#1{#2}$}}

\begin{document}

\title[Supersymmetric $t$-$J$ models with long-range interactions]{Supersymmetric $t$-$J$ models
  with long-range interactions: partition function and spectrum}

\author{B.~Basu-Mallick$^1$, N.~Bondyopadhaya$^2$, J.A.~Carrasco$^3$, F.~Finkel$^3$,
  A.~Gonz\'alez-L\'opez$^3$}

\address{$^1$Theory Division, Saha Institute of Nuclear Physics, HBNI, 1/AF Bidhan Nagar, Kolkata
  700 064, India} \address{$^2$Integrated Science Education and Research Centre, Siksha-Bhavana,
  Visva-Bharati, Santiniketan 731 235, India} \address{$^3$Departamento de F\'{\i}sica
  Te\'{o}rica, Universidad Complutense de Madrid, 28040 Madrid, Spain}

\eads{\mailto{bireswar.basumallick@saha.ac.in}, \mailto{nilanjan.iserc@visva-bharati.ac.in},
  \mailto{joseacar@ucm.es}, \mailto{ffinkel@ucm.es}, \mailto{artemio@ucm.es}}
\date{\today}
\begin{abstract}

  We study the spectrum of the long-range supersymmetric $\su(m)$ $t$-$J$ model of Kuramoto and
  Yokoyama in the presence of an external magnetic field and a charge chemical potential. To this
  end, we first establish the precise equivalence of a large class of models of this type to a
  family of $\su(1|m)$ spin chains with long-range exchange interactions and a suitable chemical
  potential term. We exploit this equivalence to compute in closed form the partition function of
  the long-range $t$-$J$ model, which we then relate to that of an inhomogeneous vertex model with
  simple interactions. From the structure of this partition function we are able to deduce an
  exact formula for the restricted partition function of the long-range $t$-$J$ model in subspaces
  with well-defined magnon content in terms of its analogue for the equivalent vertex model. This
  yields a complete analytical description of the spectrum in the latter subspaces, including the
  precise degeneracy of each level, by means of the supersymmetric version of Haldane's motifs and
  their related skew Young tableaux. As an application, we determine the structure of the motifs
  associated with the ground state of the spin~$1/2$ model in the thermodynamic limit in terms of
  the magnetic field strength and the charge chemical potential. This leads to a complete
  characterization of the distinct ground state phases, determined by their spin content, in terms
  of the magnetic field strength and the charge chemical potential.
\end{abstract}


\maketitle

\section{Introduction}
\label{sec.intro}

Among lattice models of strongly correlated fermions, the $t$-$J$ model holds a prominent position
due to its role in the theoretical description of high-temperature superconductivity and as an
example of a simple model featuring spin-charge separation~\cite{Sc87,ZR88,EKS92}. In this model
each lattice site can be either empty or occupied by one fermion, which interacts with its nearest
neighbors through spin exchange and charge repulsion and can also hop between contiguous lattice
sites.. In the one-dimensional case the $t$-$J$ model is of particular interest, as it is both
supersymmetric and exactly solvable through the nested Bethe ansatz when its parameters are
suitably related~\cite{Sc87,Su75,BB90,KY90,Sa91,EK92}.
In the early nineties, Kuramoto and Yokoyama~\cite{KY91,Ku93} introduced a long-range version of
the supersymmetric $t$-$J$ model featuring $1/r^2$ interactions, which reduces to the $\su(2)$
Haldane--Shastry (HS) chain \cite{Ha88,Sh88} in the high density limit (i.e., when all the sites
are occupied). Among other interesting features, the Kuramoto--Yokoyama (KY) model exhibits strong
spin-charge separation, in the sense that at low temperatures the spin and charge velocities are
respectively independent of the charge density and the magnetization.
At low energies the KY model is known to be a Luttinger liquid~\cite{KY91b}, with spin and charge
excitations independently described by a $c=1$ conformal field theory (CFT).

The supersymmetric KY model is exactly solvable through the asymptotic Bethe ansatz pioneered by
Sutherland and Shastry~\cite{SS93}, as its energies can be found in principle by solving a system
of transcendental equations in the asymptotic momenta~\cite{Ka92,Ka92b}. However, this method does
not completely determine the spectrum, since it does not provide complete information on the
degeneracy of each level. Based on numerical calculations, Wang et al.~\cite{WLC92} proposed an
empirical description of the degeneracies of the spectrum of the $\su(2)$ KY model reminiscent of
the rule for filling the border strips associated to Haldane's motifs~\cite{HHTBP92,Ha93,BBHS07}.
This description, however, is known to be incorrect in certain situations, although the needed
corrections vanish in the thermodynamic limit~\cite{WLC92}. Inspired by the equivalence between
the $\su(2)$ KY model and the $\su(1|2)$ supersymmetric HS chain (up to a term proportional to the
total electric charge), Saiga and Kuramoto~\cite{SK99} conjectured a description of the former
model's spectrum essentially in terms of $\su(1|2)$-supersymmetric Haldane motifs, which accounted
for the numerical results for $N\le 16$ spins. To the best of our knowledge, this conjecture has
remained unproved in the literature.

In this paper we address and completely solve the problem of finding a full description of the
spectrum of the supersymmetric $\su(m)$ KY model with a general chemical potential term for a
finite number of sites, including the determination of the levels' degeneracies and spin content.
In particular, our results provide a rigorous proof of the Saiga--Kuramoto conjecture for
arbitrary $m$ and $N$. Our approach, which we shall now briefly summarize, is new and bypasses the
usual machinery of the asymptotic Bethe ansatz, transfer matrix, Yangian highest-weight states,
etc. Indeed, we start by establishing the precise connection between the $\su(m)$ KY model and the
$\su(1|m)$ supersymmetric HS spin chain with a suitable chemical potential term for arbitrary $m$,
thus generalizing the well-known result for $m=2$. Since the partition function of the latter
chain was recently evaluated in Ref.~\cite{FGLR18} by using Polychronakos's freezing trick, this
immediately yields a novel closed formula for the partition function of the $\su(m)$ KY model. A
remarkable property of this partition function is that it can be recast as the partition function
of an equivalent (inhomogeneous) vertex model with simple interactions~\cite{BBH10}. We show that
both partition functions are polynomials in appropriate variables, whose coefficients are nothing
but the corresponding restricted partition functions on subspaces with well-defined magnon
content. This crucial observation provides a closed formula for the restricted partition function
of the $\su(m)$ KY model (in the presence of an external magnetic field and a chemical potential
term) on each of these subspaces. Finally, by analyzing the restricted partition function of the
equivalent vertex model we are able to express the spectrum of the $\su(m)$ KY model in each
subspace with well-defined magnon content in terms of suitably restricted supersymmetric Haldane
motifs and their corresponding Young tableaux.
This yields a complete and rigorous description of the spectrum in the latter subspaces, including
a systematic way for determining the degeneracy of each level, which implies the Saiga--Kuramoto
conjecture as a particular case.

It should be noted that, while the traditional freezing trick allows one to compute the partition
function (and, in principle, the spectrum) of certain integrable systems, it does not provide any
information about the corresponding wave functions. On the other hand, the analysis of the
spectrum of the $\su(m)$ KY model by the method described above, extends the freezing trick and,
more importantly, the equivalence to a vertex model, to subspaces of the Hilbert space with
well-defined magnon content. It is in fact this connection with a restricted vertex model what
makes it possible to identify all energy eigenvalues within any subspace of the Hilbert space with
well-defined magnon content. In other words, our approach not only yields the complete spectrum of
the $\su(m)$ KY model but also the magnon numbers or spin content of the corresponding wave
functions.

Our approach yields several additional results that we shall now briefly discuss. In the first
place, as a consequence of the general discussion of the equivalence of the $\su(m)$ KY model to
an $\su(1|m)$ supersymmetric HS chain with a suitable chemical potential term, we construct a new
model of KY type with general elliptic interactions which can be mapped to a corresponding
$\su(1|m)$ generalization of Inozemtsev's chain~\cite{In90}. This model certainly deserves further
study, since it smoothly interpolates between the standard (nearest-neighbors) $t$-$J$ model and
the (long-range) KY model. Secondly, as an application of the description of the spectrum of the
KY model in terms of supersymmetric Young tableaux, we determine the ground state of the spin
$1/2$ model in an external magnetic field in the thermodynamic limit. In particular, we give a
complete description of the different ground state phases, characterized by their spin content
---i.e., $\su(1|2)$, $\su(1|1)$ and $\su(0|2)$, apart from the trivial phases consisting of only
holes or fermions of one type--- in terms of the magnetic field strength and the charge chemical
potential. This description goes beyond previously known results, which are restricted to the
genuinely~$\su(1|2)$ phase. In particular, we show that the strong spin-charge separation
characteristic of the long-range $t$-$J$ model at low temperatures~\cite{KK95} occurs in all
nontrivial phases.

The paper is organized as follows. In Section~\ref{sec.models} we introduce the model and show its
equivalence to a supersymmetric HS chain with suitable chemical potential terms. We also introduce
the model's elliptic version and discuss its connection with the supersymmetric extension of
Inozmetsev's elliptic chain. The model's partition function is then computed in
Section~\ref{sec.PF} by exploiting its equivalence to a supersymmetric HS chain. In
Section~\ref{sec.motifs} we recast the partition function as that of a suitable inhomogeneous
vertex model and derive the model's restricted partition function on subspaces with well-defined
magnon content. As explained above, this yields an explicit and complete description of the
spectrum on the latter subspaces, including each level's degeneracy, in terms of suitable
supersymmetric Young tableaux. Section~\ref{sec.GS} is devoted to a complete analysis of the
ground state phases of the spin~$1/2$ model in an external magnetic field in the thermodynamic
limit. In Section~\ref{sec.conc} we present our conclusions and outline several future
developments. The paper ends with a short technical appendix in which we present the proof of a
new result regarding the degeneracy of reverse motifs.

\section{The models}\label{sec.models}

We shall deal in this paper with a class of~$\su(m)$ $t$-$J$ type models, consisting of a
one-dimensional lattice with $N$ sites each of which can be either empty or occupied by a single
fermion with $m$ internal degrees of freedom. We shall be mainly interested in long-ranged models
of the latter type, in which the spin and charge interactions among the fermions and their hopping
amplitude involve all possible pairs of lattice sites. More precisely, we shall take as the
model's Hamiltonian
\begin{equation}
  \fl
  H_0=\sum_{\mathclap{1\le i<j\le N}}\cP\bigg\{-t_{ij}\sum_{\si=1}^m(\dc_{i\si}c_{j\si}
  +\dc_{j\si}c_{i\si})+2J_{ij}\bT_i\cdot\bT_j-2V_{ij}n_in_j\bigg\}\cP\equiv
  \sum_{\mathclap{1\le i<j\le N}}H_{ij}\,,
  \label{tJgen}
\end{equation}
or equivalently\footnote{Here and in what follows, unless otherwise stated sums and products over
  Latin indexes run over the set~$1,\dots,N$, while Greek indices range from~$1$ to~$m$.}
\begin{equation}
  \label{tJgen2}
  H_0=\sum_{i\ne j}\cP\bigg\{-t_{ij}\sum_\si\dc_{i\si}c_{j\si}+
  J_{ij}\bT_i\cdot\bT_j-V_{ij}n_in_j\bigg\}\cP\,,
\end{equation}
where~$t_{ij}=t_{ji}$, $J_{ij}=J_{ji}$, $V_{ij}=V_{ji}$ are real constants. We shall also assume
that the model~\eqref{tJgen}-\eqref{tJgen2} is \emph{translation-invariant}, i.e., that
\begin{equation}\label{tJV}
  t_{ij}=t(i-j)\,,\quad
  J_{ij}=J(i-j)\,,\quad V_{ij}=V(i-j)
\end{equation}
with
\begin{equation}\label{trinv}
  t(x)=t(-x)=t(N-x)\,,
\end{equation}
and similarly for $J(x)$, $V(x)$. In the latter equations~$\dc_{i\si}$ (respectively~$c_{i\si}$)
denotes the operator creating (resp.~destroying) a fermion of type~$\si$ at site~$i$ and
$n_i=\sum_\si n_{i\si}$, where $n_{i\si}=\dc_{i\si}c_{i\si}$, is the total number of fermions at
site~$i$. The operator $\cP$ is the projector onto single-occupancy states, in which each site is
occupied by at most one fermion. Finally, $\bT_i\equiv(T_i^1,\dots,T_i^{m^2-1})$, where~$T_i^r$ is
the~$r$-th $\su(m)$ Hermitian generator in the fundamental representation acting on the $i$-th
site (with a suitable normalization that we shall specify below). Thus the first term
(proportional to $t_{ij}$) in Eqs.~\eqref{tJgen}-\eqref{tJgen2} accounts for the hopping of
fermions between sites~$i$ and~$j$, while the last two terms respectively model the spin
(exchange) and charge interaction between the latter sites.

The Hamiltonian~\eqref{tJgen}-\eqref{tJgen2} encompasses several well-known models, which we shall
briefly review. To begin with, note that a nearest-neighbors version of the
Hamiltonian~\eqref{tJgen}-\eqref{tJgen2} is obtained by taking~$t(x)$, $J(x)$, $V(x)$ proportional
to the $N$-periodic extension of
\begin{equation}\label{tde}
  \de_{1,x}+\de_{N-1,x}\,,\qquad 1\le x\le N-1\,.
\end{equation}
When~$m=2$, this is the original $t$-$J$ model introduced in Ref.~\cite{Sc87}. On the other hand,
the long-ranged supersymmetric Kuramoto--Yokoyama model~\cite{KY91,Ku93} follows from
Eqs.~\eqref{tJgen2}-\eqref{tJV} when
\[
  t(x)=J(x)=4V(x)=\frac{t\pi^2}{N^2}\sin^{-2}\bigl(\pi x/N\bigr)\,,
\]
where~$t$ is a positive real parameter. More generally, when~$m=2$ the
model~\eqref{tJgen2}-\eqref{tJV} is supersymmetric provided that $t(x)=J(x)=4V(x)$. In fact, one
of our aims in this section is to generalize the latter result to the $\su(m)$ case with arbitrary
$m>2$. As we shall next see, the key idea in this respect is to map the original
Hamiltonian~\eqref{tJgen} to that of a suitable supersymmetric spin chain in which the holes are
regarded as bosons.

More precisely, consider a one-dimensional lattice (\emph{spin chain}) each of whose sites are
occupied either by a boson or an~$\su(m)$ fermion. Thus the model's Hilbert space
is~$\hat\cH=\otimes_{i=1}^N\hat\cH_i$, where~$\hat\cH_i$ is the linear span of the one-particle
states $\db_i\ket{\hat\Om}_i$, $\df_{i\si}\ket{\hat\Om}_i$ with $\si=1,\dots,m$, $\db_i$,
$\df_{i\si}$ are the operators creating respectively a boson and a fermion of type~$\si$ at the
$i$-th site, and~$\ket{\hat\Om}_i$ is the vacuum. Similarly, denote by $\cH=\otimes_{i=1}^N\cH_i$
the Hilbert space of the original model~\eqref{tJgen}, where~$\cH_i$ is spanned by the
vacuum~$\ket\Om_i$ and the one-particle states~$\dc_{i\si}\ket{\Om}_i$. We now introduce the
unitary mapping~$\vp:\cH\to\hat\cH$ defined by
\[
  \vp\ket\Om_i=\db_i\ket{\hat\Om}_i\,,\quad
  \vp\big(\dc_{i\si}\ket\Om_i\big)=\df_{i\si}\ket{\hat\Om}_i\,.
\]
This mapping induces a natural way of associating to each linear operator $A$ acting on~$\cH$ a
corresponding linear operator~$\hat A=\vp A\vp^{-1}=\vp A \vp^\dagger$ acting on $\hat\cH$. Note,
in particular, that
$(A^\dagger)^{\textstyle\hat{}}=\vp A^\dagger \vp^\dagger\equiv\hat A^\dagger$. It is
straightforward to check that under this correspondence $\hat c_{i\si}=\db_if_{i\si}$, since both
operators agree on the canonical basis of~$\hat\cH_i$. Note than on~$\cH_i$ we
have~$c_{i\si}=\cP c_{i\si}\cP$, so that we can also write
\begin{equation}\label{PcP}
  (\cP c_{i\si}\cP)^{\textstyle\hat{}}=\db_if_{i\si}\equiv X_i^{0\si},
\end{equation}
and therefore (since $\cP$ is Hermitian, being a projector)
\begin{equation}\label{PdcP}
  (\cP\dc_{i\si}\cP)^{\textstyle\hat{}}=\hat c_{i\si}^\dagger=\df_{i\si}b_i\equiv X_i^{\si0}\,.
\end{equation}
We shall also need in the sequel the relations
  \begin{eqnarray}
    \smash{\big(\cP \dc_{i\si}c_{i\si'}\cP\big)}^{\textstyle\hat{}}
    &=\df_{i\si}f_{i\si'}\equiv
      X_i^{\si\si'}\,,
      \label{Xissp}\\
    {[\cP(1-n_i)\cP]}^{\,\textstyle\hat{}}&=\db_ib_i\equiv X_i^{00}\,,
                                          \label{niXi00}
  \end{eqnarray}
which easily follow from the previous ones. For instance, taking into account
that~$c_{i\si'}\cP=\cP c_{i\si'}\cP$ and~$\cP^2=\cP$ we have
\begin{eqnarray*}
  \smash{\big(\cP \dc_{i\si}c_{i\si'}\cP\big)}^{\textstyle\hat{}}
  &=\smash{\big(\cP \dc_{i\si}\cP\cdot\cP c_{i\si'}\cP\big)}^{\textstyle\hat{}}
  \\
  &=\df_{i\si}b_i\db_if_{i\si'}=X_i^{\si\si'}+\df_{i\si}\db_if_{i\si}b_i
    =X_i^{\si\si'},
\end{eqnarray*}
since~$f_{i\si}b_i=0$ on~$\hat\cH_i$.

Consider next the $\su(1|m)$ supersymmetric permutation
operators~$P_{ij}^{(1|m)}:\hat\cH\to\hat\cH$ (with~$i<j$), whose action on the canonical basis
\[
  \ket{\si_1\cdots \si_N}\equiv \da_{1\si_1}\cdots\da_{N\si_N}\ket{\hat\Om}\,,\quad
  \si_1,\dots,\si_N\in\{0,\dots,m\},
\]
where~$a_{i0}=b_i$ and~$a_{i\si}=f_{i\si}$ for~$\si\ge1$
and~$\ket{\hat\Om}=\otimes_i\ket{\hat\Om}_i$ is the global vacuum, is given by~\cite{BB06}
\[
  P_{ij}^{(1|m)}\ket{\cdots\si_i\cdots\si_j\cdots}=\ep(\bsi)\ket{\cdots\si_j\cdots\si_i\cdots}\,.
\]
The sign~$\ep(\bsi)$ is $1$ (respectively $-1$) if $\si_i=\si_j=0$ (resp.~$\si_i,\si_j\ge1$),
while for $\si_i\si_j=0$ and $\si_i\ne\si_j$ it is equal to the number of fermionic spins $\si_k$
with~$i+1\le k\le j-1$. It is well known~\cite{Ha93,HB00,BB06} that~$P_{ij}^{(1|m)}$ can be
expressed in terms of creation and annihilation operators as follows:
\[
  P_{ij}^{(1|m)}=\sum_{\al,\be=0}^m\da_{i\al}\da_{j\be}a_{i\be}a_{j\al}=
  \sum_{\al,\be=0}^m(-1)^{p(\be)}X_{i}^{\al\be}X_j^{\be\al}\,,
\]
where~$p(0)=0$ and $p(\si)=1$ for~$\si\ge1$. We thus have
\begin{equation}\label{Pij1m}
  P_{ij}^{(1|m)}=X_i^{00}X_j^{00}+\sum_\si(X_i^{\si0}X_j^{0\si}-X_i^{0\si}X_j^{\si0})
  -P_{ij}\,,
\end{equation}
where
\[
  P_{ij}=\sum_{\si,\si'}X_{i}^{\si\si'}X_j^{\si'\!\si}
\]
is the ordinary permutation operator when acting on purely fermionic states.

Our next goal is to relate the product~$\bT_i\cdot\bT_j$ appearing in the
Hamiltonian~\eqref{tJgen} with the supersymmetric permutation operator~$P_{ij}^{(1|m)}$. To this
end, note first of all that the components~$T_i^r$ of~$\bT_i$ are defined in the usual way as
\begin{equation}\label{Tir}
  T_i^r=\sum_{\si,\si'}T^r_{\si\si'}\dc_{i\si}c_{i\si'}\,,
\end{equation}
where the complex numbers~$T^r_{\si\si'}$ are the matrix elements of the~$r$-th (Hermitian)
generator of~$\su(m)$ in the fundamental representation. We shall normalize the~$m\times m$
matrices~$T^r\equiv \big(T^r_{\si\si'}\big)_{1\le\si,\si'\le m}$ so that
\[
  \tr(T^rT^s)=\frac12\,\de_{rs}\,.
\]
In particular, when~$m=2$ the operator $T^r$ can be taken as the usual spin~$1/2$
operator~$S^r=\si^r/2$, where~$\si^r$ is the $r$-th Pauli matrix. In order to
relate~$\bT_i\cdot\bT_j$ with~$P_{ij}^{(1|m)}$ we shall make use of the identity
\begin{equation}\label{comp}
  2\sum_{r=1}^{m^2-1}(T^r)_{\si\si'}(T^r)_{\mu\mu'}=\de_{\si\mu'}\de_{\si'\mu}
  -\frac1m\,\de_{\si\si'}\de_{\mu\mu'}\,,
\end{equation}
which is a direct consequence of the completeness of the generators $T^r$ together with the
identity matrix. From Eqs.~\eqref{Xissp}, \eqref{Tir} and~\eqref{comp} we obtain
\begin{eqnarray}
  \fl
  2(\cP\bT_i\cdot\bT_j\cP)^{\textstyle\hat{}}
  &=2\sum_{r=1}^{m^2-1}\sum_{\substack{\si,\si'\cr\mu,\mu'}}T^r_{\si\si'}T^r_{\mu\mu'}
    X_i^{\si\si'}
    X_j^{\mu\mu'}=\sum_{\si,\si'}X_i^{\si\si'}X_j^{\si'\!\si}-\frac1m\sum_{\si,\mu}X_i^{\si\si}X_i^{\mu\mu}
    \nonumber\\
  \fl
  &= P_{ij}-\frac1m\,\hat n_i\hat n_j\,,
    \label{TiTjPij}
\end{eqnarray}
where~$\hat n_k=\sum_\si \df_{k\si}f_{k\si}$ denotes the total number of fermions (created by the
operators $\df_{k\si}$) at the~$k$-th site. From Eq.~\eqref{Pij1m} for the supersymmetric
permutation operator~$P_{ij}^{(1|m)}$ we obtain, after some algebra,
\begin{equation}\label{PijAij}
  P_{ij}^{(1|m)}+\hat n_i+\hat n_j-1=\hat A_{ij}\,,
\end{equation}
where
\[
  A_{ij}=\cP\bigg[\sum_\si(\dc_{i\si}c_{j\si}+\dc_{j\si}c_{i\si})-2\bT_i\cdot\bT_j
  +\bigg(1-\frac1m\bigg)n_in_j\bigg]\cP.
\]
Comparing with Eq.~\eqref{tJgen} we deduce that~$H_{ij}$ will be proportional to~$A_{ij}$ provided
that
\begin{equation}
  \label{SUSY}
  t(x)=J(x)=2\bigg(1-\frac1m\bigg)^{\!\!-1}V(x)\,,
\end{equation}
and in that case
\[
  \hat H_{ij}=t_{ij}\big(1-P_{ij}^{(1|m)}-\hat n_i-\hat n_j\big)\,.
\]
In other words, when condition~\eqref{SUSY} is satisfied, i.e., when~$H_0$ is of the form
\begin{equation}
  H_0=\sum_{i<j}t_{ij}\cP\bigg[-\sum_{\si}(\dc_{i\si}c_{j\si}+\dc_{j\si}c_{i\si})
  +2\bT_i\cdot\bT_j-\left(1-\tfrac1m\right)n_in_j\bigg]\cP\,,
  \label{tJSUSY}
\end{equation}
the corresponding Hamiltonian~$\hat H_0$ is given by
\[
  \hat H_0=\sum_{i<j}t_{ij}(1-P_{ij}^{(1|m)}-\hat n_i-\hat n_j)\\
  =\sum_{i<j}t_{ij}(1-P_{ij}^{(1|m)})-\sum_{i\ne j}t_{ij}\hat n_j\,.
\]
Note that so far we have not used the translation-invariance conditions~\eqref{tJV}-\eqref{trinv},
so that the previous result is valid in full generality. On the other hand, when the
model~\eqref{tJgen} is translation-invariant we can use Eqs.~\eqref{tJV}-\eqref{trinv} to further
simplify the last term in the previous equation. Indeed, in this case
\[
  \sum_{i\ne j}t_{ij}\hat n_j=\sum_j\bigg(\sum_{i;i\ne j}t_{ij}\bigg)\hat n_j\,,
\]
with
\[
  \fl
  \sum_{i;i\ne j}t_{ij}=\sum_{k=1-j}^{-1}t(k)+\sum_{k=1}^{N-j}t(k)
  = \sum_{k=N-j+1}^{N-1}t(k-N)+\sum_{k=1}^{N-j}t(k) = \sum_{k=1}^{N-1}t(k)\equiv t_0\,,
\]
so that
\begin{equation}
  \hat H_0
  =\sum_{i<j}t_{ij}(1-P_{ij}^{(1|m)})-t_0\cF,
  \label{hatH}
\end{equation}
where
\[
  \cF\equiv\sum_{i}\hat n_{i}
\]
is the total number of fermions.

Summarizing, we have shown that the translation-invariant $\su(m)$ model~\eqref{tJgen}-\eqref{tJV}
is supersymmetric if its coefficients are related by Eq.~\eqref{SUSY}. When this is the case this
general model reduces to~\eqref{tJSUSY}, which is equivalent to the~$\su(1|m)$ supersymmetric spin
chain~\eqref{hatH}. It should be stressed that the coefficient of the charge interaction
term~$n_in_j$ in the supersymmetric $t$-$J$ Hamiltonian~\eqref{tJSUSY} must depend on~$m$ as
specified in the latter equation, a fact that does not seem to have been previously noted in the
literature.

Note that Eqs.~\eqref{tJSUSY}--\eqref{hatH} are also valid for~$m=1$. In this case the
Hamiltonian~\eqref{tJSUSY} is that of a free fermion system (the terms proportional
to~$\bT_i\cdot\bT_j$ and $n_in_j$ vanish identically), as first noted by Göhmann and
Wadati~\cite{GW95}. This fact was recently exploited in Refs.~\cite{CFGRT16,CFGR17} to evaluate
the entanglement entropy of the ground state of (translation-invariant) $\su(1|1)$ spin chains of
HS type.

In the rest of this work we shall be mainly concerned with the supersymmetric $t$-$J$
model~\eqref{tJSUSY} with
\begin{equation}\label{tKY}
  t(x)=t(\pi/N)^2\sin^{-2}(\pi x/N)\,,
\end{equation}
which is the~$\su(m)$ version of the original KY model. In this case~\cite{CP78,FG05}
\begin{equation}
  \label{sinsum}
  \sum_{i\ne j}\sin^{-2}\bigl(\pi(i-j)/N\bigr)=\frac N3\,(N^2-1)\,,
\end{equation}
so that
\begin{equation}\label{t0HS}
  t_0=\frac{t\pi^2}{3N^2}(N^2-1)\,.
\end{equation}
Hence the Hamiltonian~$\hat H_0$ of the equivalent $\su(1|m)$ supersymmetric spin chain can be
written as
\begin{equation}
  \label{HtJHS}
  \hat H_0=\frac{2t\pi^2}{N^2}\bigg[ H_{\mathrm{HS}}^{(1|m)}-
  \frac 16(N^2-1)\cF\bigg]\,,
\end{equation}
where
\begin{equation}
  \label{HHS}
  H_{\mathrm{HS}}^{(1|m)}=\frac12\sum_{i<j}\frac{1-P_{ij}^{(1|m)}}{\sin^2(\pi(i-j)/N)}
\end{equation}
is the Hamiltonian of the $\su(1|m)$ Haldane--Shastry spin chain~\cite{Ha88,Sh88,Ha93,HH94}. In
fact, the model~\eqref{HtJHS} was introduced by Kawakami in the early 90's~\cite{Ka92}. On the
other hand, when~$t(x)$ is proportional to Eq.~\eqref{tde} the Hamiltonian~\eqref{tJSUSY} reduces
to the $\su(m)$ version of the original (nearest-neighbor) \hbox{$t$-$J$} model
\begin{equation}
  \fl
  H_0=t\sum_{i}\cP\bigg[-\sum_{\si}(\dc_{i\si}c_{i+1,\si}+\dc_{i+1,\si}c_{i\si})
  +2\bT_i\cdot\bT_{i+1}-\left(1-\tfrac1m\right)n_in_{i+1}\bigg]\cP,
  \label{tJSUSYnn}
\end{equation}
where $N+1\equiv1$. The equivalent $\su(1|m)$ supersymmetric chain Hamiltonian is given by
\begin{equation}\label{US}
  \hat H_0=t\sum_i\big(1-P_{i,i+1}^{(1|m)}\big)-2t\cF\,,
\end{equation}
where $P_{N,N+1}^{(1|m)}\equiv P_{1N}^{(1|m)}$, which is essentially the $\su(1|m)$
Uimin--Lai--Sutherland model~\cite{Ui70,La74,Su75}.

Next, inspired by Inozemtsev's elliptic spin chain~\cite{In90}, we introduce a one-parameter
family of supersymmetric $\su(m)$ $t$-$J$ models~\eqref{tJSUSY} which smoothly interpolate between
the $\su(m)$ KY model~\eqref{tJSUSY}-\eqref{tKY} and the (periodic) nearest-neighbors $\su(m)$
$t$-$J$ model~\eqref{tJSUSYnn}. More precisely, let
\begin{equation}\label{ellt}
  t(x)=t\bigg(\frac\al\pi\bigg)^{\!\!2}\!\sinh^2(\pi/\al)\bigg(\wp_N(x)
  -\frac{2\tilde\eta_1}{\al^2}\bigg)\,,
\end{equation}
where
\[
  \wp_N(x)\equiv\wp(x;N/2,\iu\al/2),\quad \tilde\eta_1\equiv\ze(1/2;1/2,\iu N/(2\al))
\]
and~$\al>0$. In the latter formulas~$\wp(x;\om_1,\om_3)$ and $\ze(x;\om_1,\om_3)$ denote
respectively the Weierstrass elliptic function with half-periods $\om_1,\om_3$ and its
corresponding zeta function, defined by
\begin{eqnarray*}
  \ze(z;\om_1,\om_3)&=\frac1z+\sum_{l,n\in\ZZ^2-\{(0,0)\}}\bigg[\frac1{z-2l\om_1-2n\om_3}
                      +\frac{z}{(2l\om_1+2n\om_3)^2}\bigg]\,,\\
  \wp(z;\om_1,\om_3)&=-\ze'(z;\om_1,\om_3)\,.
\end{eqnarray*}
It can be shown~\cite{In05,FG14JSTAT} that when~$1\le x\le N-1$
\[
  \lim_{\al\to0+}t(x)=t(\de_{1,x}+\de_{-1,x}),
\]
while the~$\al\to+\infty$ limit of Eq.~\eqref{ellt} is Eq.~\eqref{tKY}. Note that the constant
$t_0$ for the function~\eqref{ellt} is given by
\[
  t_0=\frac{2t}{\pi^2}\sinh^2(\pi/\al)(\tilde\eta_1-\eta_1)\,,\quad
  \eta_1\equiv\ze(1/2;1/2,\iu/(2\al))\
\]
(see, e.g., Ref.~\cite{FG14}). Thus the Hamiltonian~\eqref{hatH} of the~$\su(1|m)$ supersymmetric
spin chain equivalent to the~$\su(m)$ $t$-$J$ model~\eqref{tJSUSY} with elliptic
interactions~\eqref{ellt} is given by
\[
  \hat H_0 = tH_{I}^{(1|m)}-\frac{2t}{\pi^2}\,\sinh^2(\pi/\al)(\tilde\eta_1-\eta_1)\cF,
\]
where
\[
  H_{I}^{(1|m)}=\bigg(\frac\al\pi\bigg)^{\!\!2}\!\sinh^2(\pi/\al)
  \sum_{i<j}\bigg(\wp_N(i-j)-\frac{2\tilde\eta_1}{\al^2}\bigg)(1-P_{ij}^{(1|m)})
\]
is the~$\su(1|m)$ version of Inozemtsev's elliptic spin chain~\cite{In90}. In fact, for $m=1$ the
partition function and thermodynamics of the latter chain were derived in Ref.~\cite{FG14JSTAT},
and the entanglement entropy of its ground state was analyzed in Ref.~\cite{CFGRT16}.

\section{Partition function}\label{sec.PF}

In this section we shall compute in closed form the partition function of the~$\su(m)$
KY model~\eqref{tJSUSY}-\eqref{tKY} by exploiting its equivalence with the
$\su(1|m)$ spin chain Hamiltonian of Haldane--Shastry type~\eqref{HtJHS}-\eqref{HHS}. As a matter
of fact, we shall consider the more general Hamiltonian
\begin{equation}
  \label{HKYmu}
  H=H_0-\frac12\sum_{\si=1}^{m-1}h_\si(n^\si-n^m)-\mu_c\sum_\si n^\si\equiv H_0+H_1,
\end{equation}
where $H_0$ is given by Eqs.~\eqref{tJSUSY}-\eqref{tKY} and
\[
  n^\si\equiv\sum_in_{i\si}
\]
denotes the total number of fermions of type~$\si$. The last term in~$H_1$ is the chemical
potential of the fermions (or, equivalently, of the total electric charge), while the first one
has a natural interpretation as arising from the interaction with an external $\su(m)$ magnetic
field with strengths $h_1,\dots,h_{m-1}$. Indeed, for $m=2$ the term $-(h_1/2)(n^1-n^2)$ equals
$-h_1S^z$, where~$S^z$ is the $z$ component of the total spin operator. This is indeed the
contribution to the energy arising from the interaction with the magnetic field $h_1\mathbf e_z$
of a charged fermion (with gyromagnetic ratio~$g=2$, and unit mass and electric charge in natural
units). More generally, for arbitrary $m\ge2$ we have
\[
  n^\si-n^m=\sum_i(n_{i\si}-n_{im}),
\]
where the operators $\iu(n_{k\si}-n_{km})$ generate the standard $\su(m)$ Cartan subalgebra at the
$k$-th site. By Eq.~\eqref{Xissp}, the $\su(1|m)$ spin chain Hamiltonian~$\hat H$ equivalent to
$H$ is~$\hat H=\hat H_0+\hat H_1$, where
\[
  \hat H_1=-\frac12\sum_{\si=1}^{m-1}h_\si(\cN_\si-\cN_m)-\mu_c\cF
\]
and~$\cN_\si\equiv\sum_i\hat n^\si$ is the total numbers of fermions (created
by~$\df_{i\si}f_{i\si}$) of type~$\si$. More explicitly, we have
\begin{equation}
  \label{hatH0H1}
  \hat H=J H_{\mathrm{HS}}^{(1|m)}-\frac12\sum_{\si=1}^{m-1}h_\si(\cN_\si-\cN_m)-
  (t_0+\mu_c)\cF\,,
\end{equation}
with $t_0$ given by Eq.~\eqref{t0HS} and $J=2t\pi^2/N^2$. The latter equation can be more
concisely rewritten as
\begin{equation}
  \label{hatHmu}
  \hat H=J H_{\mathrm{HS}}^{(1|m)}-\sum_\si\mu_\si\cN_\si,
\end{equation}
where $\mu_\si$ is the chemical potential of the fermion of type~$\si$, given by
  \begin{eqnarray}\label{musi}
    \mu_\si&=\frac12\,h_\si+\mu_c+t_0\,,\quad 1\le\si\le m-1\,;\\
    \mu_m&=-\frac12\,\sum_{\si=1}^{m-1}h_\si+\mu_c+t_0\,.
           \label{musim}
  \end{eqnarray}
  From the above remarks, it follows that the partition function~$\cZ$ of the supersymmetric
  $\su(m)$ KY model~\eqref{HKYmu} coincides with that of the $\su(1|m)$ supersymmetric
  Haldane--Shastry chain with a chemical potential term given in Eq.~\eqref{hatHmu}. The partition
  function of the latter model has been recently evaluated in Ref.~\cite{FGLR18} by taking
  advantage of its connection with the supersymmetric spin Sutherland model via Polychronakos's
  freezing trick~\cite{Po93,Po94}, with the result
\begin{equation}\label{PF}
  \cZ(q;\bmu)=\sum_{\mathclap{\bk\in\cP_N}}d(\bk)q^{\sum\limits_{i=1}^{r-1}JK_i(N-K_i)}
  \prod_{i=1}^{\mathclap{N-r}}\big(1-q^{JK'_i(N-K'_i)}\big)\,.
\end{equation}
Here~$q\equiv\e^{-1/T}$, $\bk=(k_1,\dots,k_r)\in(\NN\cup\{0\})^r$, $\cP_N$ is the set of
partitions (with order taken into account) of the integer~$N$, $K_i\equiv\sum_{j=1}^ik_j$, and
the~$N-r$ positive integers~$K'_i$ are defined by
\[
  \{K_1',\dots,K'_{N-r}\}=\{1,\dots,N-1\}-\{K_1,\dots,K_{r-1}\}\,.
\]
For each multiindex~$\bk\in(\NN\cup\{0\})^r$, the coefficient~$d(\bk)$ is defined by
\begin{equation}\label{dbk}
  d(\bk)=\prod_{i=1}^r\sum_{j=0}^{\min(k_i,m)}e_j(q^{-\bmu})\,,\quad
  q^{-\bmu}\equiv(q^{-\mu_1},\dots,q^{-\mu_m}),
\end{equation}
where
\begin{equation}\label{ejdef}
  e_j(y_1,\dots,y_m)\equiv\en\sum_{\mathclap{1\le i_1<\cdots<i_j\le m}}\en y_{i_1}\cdots y_{i_j}
\end{equation}
denotes the elementary symmetric polynomial of degree~$j\le m$ in $m$ variables~$y_1,\dots,y_m$
(with~$e_0\equiv1$). By the above remarks, Eq.~\eqref{PF} yields also the partition function of
the supersymmetric~$\su(m)$ KY model~\eqref{HKYmu}. This is one of the main results of the present
work. 

Equation~\eqref{dbk} for $d(\bk)$ can be considerably simplified introducing the
numbers~$\nu_i(\bk)$ defined by
\begin{eqnarray*}
  \nu_l(\bk)&=\left|\big\{i\in\{1,\dots,r\}:k_i=l\big\}\right|,\quad l=1,\dots,m-1\,,\\
  \nu_{m}(\bk)&=\left|\big\{i\in\{1,\dots,r\}:k_i\ge m\big\}\right|,
\end{eqnarray*}
where~$|A|$ denotes the cardinal of the set~$A$. We then have
\begin{equation}
  \label{degm}
  d(\bk)=\prod_{l=1}^{m-1}\Bigg[\sum_{j=0}^le_j(q^{-\bmu})\Bigg]^{\nu_l(\bk)}\cdot
  \prod_{j=1}^m(1+q^{-\mu_j})^{\nu_m(\bk)}.
\end{equation}
For instance, in the $\su(2)$ and~$\su(3)$ cases we respectively have
\[
  d(\bk)=(1+q^{-\mu_1}+q^{-\mu_2})^{\nu_1(\bk)}[(1+q^{-\mu_1})(1+q^{-\mu_2})]^{\nu_2(\bk)}
\]
and
\[
  \fl
  d(\bk)=\Big(1+{\textstyle\sum\limits_\si}\,q^{-\mu_\si}\Big)^{\nu_1(\bk)}
  \Big(1+{\textstyle\sum\limits_\si}\,q^{-\mu_\si}
  +{\textstyle\sum\limits_{\si<\si'}}q^{-(\mu_\si+\mu_{\si'})}\Big)^{\nu_2(\bk)}
  \prod_\si(1+q^{-\mu_\si})^{\nu_3(\bk)}\,.
\]
The latter expressions becomes even simpler when applied to the ``pure'' supersymmetric KY model
(i.e., without magnetic field or chemical potential terms), for which~$\mu_\si=t_0$ for
all~$\si=1,\dots,m$ according to Eqs.~\eqref{musi}-\eqref{musim}. Indeed, in this case we have
\[
  e_j(q^{-\bmu})=e_j(q^{-t_0},\dots,q^{-t_0})=\sum_{\mathclap{1\le i_1<\cdots<i_j\le m}}\en
  q^{-jt_0}=\binom mj q^{-jt_0}\,,
\]
and hence
\[
  d(\bk)=\big(1+q^{-t_0}\big)^{m\nu_m(\bk)}\prod_{l=1}^{m-1}\Bigg[\sum_{j=0}^l
  \binom mj q^{-jt_0}\Bigg]^{\nu_l(\bk)}.
\]
For instance, the partition function of the original ($\su(2)$) supersymmetric KY model is given
by
\[
  \fl
  \cZ_N(q)=\sum_{\mathclap{\bk\in\cP_N}}\big(1+2q^{-t_0}\big)^{\nu_1(\bk)}\big(1+q^{-t_0}\big)^{2\nu_2(\bk)}
  q^{\sum\limits_{i=1}^{r-1}JK_i(N-K_i)}
  \prod_{i=1}^{\mathclap{N-r}}\big(1-q^{JK'_i(N-K'_i)}\big)\,.
\]

%

\section{Spectrum and motifs}\label{sec.motifs}

In this section we shall give a complete description of the spectrum of the supersymmetric
$\su(m)$ KY model~\eqref{tJSUSY}-\eqref{tKY} ---or, more generally, the
Hamiltonian~\eqref{HKYmu}--- in each subspace with well-defined spin content in terms of the
supersymmetric version of Haldane's motifs~\cite{Ha93} and their associated skew Young
tableaux~\cite{BBHS07,BBH10,FG15}. In particular, this description implies the validity of the
Saiga--Kuramoto conjecture, which is one of the main results of this paper.

We start by recalling that the partition function~\eqref{PF} of the~$\su(1|m)$ spin
chain~\eqref{hatH0H1} ---and, hence, of the~supersymmetric $\su(m)$ KY model~\eqref{HKYmu}---
exactly coincides with the partition function of the inhomogeneous vertex model with
energies~\cite{BBH10,FGLR18}
\begin{equation}\label{specH}
  E(\bbs)=J\sum_{i=1}^{N-1}\de(s_i,s_{i+1})i(N-i)
  -\sum_{i}\mu_{s_i}\,,
\end{equation}
where~$\mu_0\equiv0$, $\bbs\in\{0,\dots,m\}^N$ and~$\de(s,s')$ is defined by
\[
  \de(s,s')=\cases{1,&$s>s' \en\text{or}\en s=s'>0$\\
    0,&$s<s'\en\text{or}\en s=s'=0$\,.
  }
\]
The first sum in Eq.~\eqref{specH} can be interpreted as the energy of a one-dimensional vertex
model with~$N+1$ vertices~$0,\dots,N$ joined by~$N$ bonds with
values~$s_1,\dots,s_N\in\{0,\dots,m\}$, the energy associated to the $i$-th vertex~being equal
to~$\de(s_i,s_{i+1})i(N-i)$. For this reason, we shall henceforth refer to the vector~$\bbs$ as
the \emph{bond vector}. Likewise, the vectors~$\bde(\bbs)$ with components~$\de(s_i,s_{i+1})$
$(1\le i\le N-1)$ in Eq.~\eqref{specH} can be identified with $\su(1|m)$
motifs~\cite{Ha93,HB00,BBH10}. Thus the spectrum of the $\su(m)$ KY model (with the correct
degeneracy for each level) can be computed from Eq.~\eqref{specH} by letting~$\bbs$ run over all
possible~$(m+1)^N$ bond vectors. It is important to note that the energies~\eqref{specH} depend not
only on the motif~$\bde$ but also on the chemical potentials~$\mu_\al$ through the last term. This
term will in general break the huge degeneracy associated to the motifs~$\bde$, which is in part
due to the invariance of the model~\eqref{hatHmu} with $\mu_\al=0$ (i.e., the $\su(1|m)$
supersymmetric HS chain) under the Yangian~$Y(\gl(1|m))$~\cite{FG15}. In other words, the general
model~\eqref{hatHmu} should be far less degenerate than the $\su(1|m)$ supersymmetric HS spin
chain. 

Let us denote by $\psi_\bbs$ the unique eigenfunction of the~supersymmetric $\su(m)$ KY
model~\eqref{HKYmu} corresponding to the eigenvalue $E(\bbs)$ \eqref{specH} associated with the
bond vector~$\bbs$. Our aim is to determine the magnon numbers or spin content of the
eigenfunction $\psi_\bbs$ directly from the structure of its bond vector~$\bbs$. To this end, we
shall now show how the spectrum of the restriction of the Hamiltonian~\eqref{HKYmu} to each
subspace with well-defined spin content can be fully generated from the motif
formula~\eqref{specH} by suitably restricting the components of the bond vector~$\bbs$.

More precisely, let us denote by
\[
  n(N_0,\dots,N_m)\equiv n(\bN)\,,\qquad N_0+\cdots+N_m\equiv|\bN|=N,
\]
the subspace of the Hilbert space of the~$\su(m)$ KY model in which each number operator~$n^{\al}$
has a well-defined value $N_\al$. We shall henceforth refer to the vector~$\bN$ as the
\emph{magnon content} of the subspace~$n(\bN)$. Note next that, by Eqs.~\eqref{HKYmu}
and~\eqref{hatH0H1}, the KY Hamiltonian~$H$ in Eq.~\eqref{HKYmu} can be written as
\begin{equation}
  H= {H}_0'-\sum_\sigma \mu_\sigma n^\sigma \,,\qquad
  H_0' \equiv H_0+t_0\sum_\si n^\si\,,
 \label{HKYmu1}
\end{equation}
where $ H_0$ and $t_0$ are respectively given by Eqs.~\eqref{tJSUSY}-\eqref{tKY} and~\eqref{t0HS}.
Clearly, $H_0'$ does not depend on the chemical potentials~$\mu_\si$ (since~$H_0$ is also
independent of~$\mu_\si$); in fact, $\widehat {H_0'} =J H_{\mathrm{HS}}^{(1|m)}$.
%
It should be noted at this point that the subspaces $n(\bN)$ are invariant under both $H$ and
${H}_0'$, since the number operators $n^\si$ obviously commute with the
operators~$\dc_{k\si}c_{l\si}$, $n_k$, $\bT_k\cdot\bT_l$.
Consequently, by Eq.~\eqref{HKYmu1}, the partition function of the Hamiltonian~$H$ can be written
as
\begin{equation}\label{Z0bN}
  \cZ(q;\bmu)=\sum_{\bN;|\bN|=N}q^{-\sum_\si \mu_\si N_\si}\cZ_0^{\bN}(q),
\end{equation}
where~$\cZ_0^{\bN}(q)$ denotes the partition function of the restriction of~${H}_0'$ to the
subspace~$n(\bN)$. We also know that
\begin{equation}\label{equiv}
  \cZ(q;\bmu)=\cZ_V(q;\bmu)\,,
\end{equation}
where~$\cZ_V(q;\bmu)$ denotes the partition function of the vertex model~\eqref{specH}. Let us now
rewrite the latter equation as
\[
  E(\bbs)=J\sum_{i=1}^{N-1}\de(s_i,s_{i+1})i(N-i) -\sum_{\si}\mu_\si N_{\si}(\bbs),
\]
where~$N_\si(\bbs)$ denotes the number of components of the bond vector~$\bbs$ equal to~$\si$.
We then have
\begin{equation}\label{ZNsi}
  \fl
  \cZ_V(q;\bmu)=\sum_{\bbs}q^{E(\bbs)}=\sum_{\mathclap{\bN;|\bN|=N}}q^{-\sum_\si\mu_\si
    N_\si}\sum_{\mathclap{\bbs;N_\al(\bbs)=N_\al}}q^{J\sum_{i=1}^{N-1}\de(s_i,s_{i+1})i(N-i)},
\end{equation}
where the last sum is restricted to all bond vectors~$\bbs$ such that~$N_\al(\bbs)=N_\al$ for
all~$\al=0,\dots,m$. Note that the right-hand sides of Eqs.~\eqref{Z0bN} and~\eqref{ZNsi} are both
polynomials in the variables~$x_\al\equiv\e^{\be\mu_\al}$ ($\al=0,\dots,m$).
Equating the coefficient of~$x_0^{N_0}\cdots x_m^{N_m}$ in both equivalent expressions
for~$\cZ(q;\bmu)$ we immediately deduce that the partition function of the restriction of $H_0'$
to the subspace~$n(\bN)$ is given by
\begin{equation}
  \label{Z0motifs}
  \cZ_0^{\bN}(q)=\sum_{\bbs:N_\al(\bbs)=N_\al}q^{J\sum_{i=1}^{N-1}\de(s_i,s_{i+1})i(N-i)}\,.
\end{equation}
It follows from the latter equation that the spectrum of~$H_0'$ in the subspace~$n(\bN)$ can be
generated from the motif formula
\[
  \fl E_0(\bbs;\bN)=J\sum_{i=1}^{N-1}\de(s_i,s_{i+1})i(N-i),\qquad \text{with}\en
  N_\al(\bbs)=N_\al\en\text{for all }\al=0,\dots,m\,.
\]
In view of Eq.~\eqref{HKYmu1}, the spectrum of the full Hamiltonian~$H$ restricted to the
subspace~$n(\bN)$ (with the correct degeneracies for all levels) is generated by the analogous
formula
\begin{equation}
  \label{specHcNmotifs}
  \fl
  E(\bbs;\bN)=E_0(\bbs;\bN)-\sum_\si\mu_\si N_\si,\qquad
  \text{with}\en N_\al(\bbs)=N_\al\en\text{for all }\al=0,\dots,m\,.
\end{equation}
We thus conclude that the spectrum of~$H$ on~$n(\bN)$ is obtained from Eq.~\eqref{specH} by
imposing the natural conditions~$N_\al(\bbs)=N_\al$ on the bond vector~$\bbs$. It follows from
this assertion that we can label the eigenfunctions of~$H$ in such a way that each $\psi_\bbs$
belongs to the subspace~$n(\bN(\bbs))$ containing exactly~$N_\al(\bbs)$ magnons of type~$\al$.
This is in fact one of the main results of this section, whose consequences we shall explore next.

To begin with, we shall use the method of Ref.~\cite{BBH10} to express the
spectrum~\eqref{specHcNmotifs} of the supersymmetric KY model~\eqref{HKYmu} on the invariant
subspace $n(\bN)$ in an alternative way. To this end, we first note that the numerical values of
the energies can be computed from the formula
\begin{equation}\label{EdeN}
  E_{\bde,\bN}=J\sum_{i=1}^{N-1}\de_i\cdot i(N-i)-\sum_\si\mu_\si N_\si\,,
\end{equation}
where now~$\bde\equiv(\de_1,\dots,\de_{N-1})$ is a supersymmetric motif, i.e., a sequence of~$N-1$
zeros and ones. Secondly, the degeneracy of each energy~$E_{\bde,\bN}$ (which could be zero) is
evaluated by counting the number of ways of filling the border strip associated to the
motif~$\bde$ according to the usual~$\su(1|m)$ rules~\cite{KKN97,FG15}, with the additional
restriction that each number~$\al\in\{0,\dots,m\}$ must appear exactly~$N_\al$ times. More
precisely, given the motif $(\de_1,\dots,\de_{N-1})$ its associated border strip is constructed by
starting with one box, and then reading the motif from left to right and adding a box below
(resp.~to the left of) the $i$-th box provided that $\de_i$ is equal to $0$ (resp.~$1$); see,
e.g., Fig.~\ref{fig.yd}. This border strip should then be filled with the numbers $0,1,\dots,m$
according to the following rules:
\begin{enumerate}[i)]
\item The numbers in each row form a nondecreasing sequence, allowing only the repetition of
  positive numbers.
\item The numbers in each column (read from top to bottom) form a nondecreasing sequence,
  allowing only the repetition of $0$.
\item Each number~$\al\in\{0,\dots,m\}$ must appear exactly~$N_\al$ times.
\end{enumerate}
Each filling of a border strip according to the previous rules is called a (skew) Young tableau.
Given such a tableau, it is straightforward to check that its associated motif can be obtained
from the bond vector~$(s_1,\dots,s_N)$ whose components are the numbers in the tableau read from
top to bottom by setting~$\de_i=\de(s_i,s_{i+1})$. The equivalence between the description of the
spectrum through Eq.~\eqref{EdeN} (where the degeneracy of a motif~$\bde$ is evaluated counting
all the fillings of its associated border strip allowed by rules i)--iii)) and
Eq.~\eqref{specHcNmotifs} essentially follows from this observation. Note also that, by
Eq.~\eqref{EdeN}, all the Young tableaux associated to a given motif have the same energy within
each invariant subspace~$n(\bN)$ .
\begin{figure}[t]
  \hfill\small
\[
  \ytableaushort{\none0,01,0,2
}
\qquad
  \ytableaushort{\none0,02,0,1
}
\qquad
  \ytableaushort{\none1,02,0,0
}
\]
\hfill
\caption{Allowed Young tableaux for the motif~$(0,1,0,0)$ in the subspace~$n(3,1,1)$.
    \label{fig.yd}}
\end{figure}

Consider, for instance, the motif~$\bde=(0,1,0,0)$ for $N=5$ particles in the~$\su(1|2)$ case. As
explained above, the degeneracy associated to this motif in each invariant subspace~$n(\bN)$ is
given by all possible Young tableaux generated from it according to the above three rules. For
instance, for the subspace~$n(3,1,1)$ it is easy to check that there are exactly three allowed
Young tableaux for the above motif (cf.~Fig.~\ref{fig.yd}), with energy $6J-\mu_1-\mu_2$. In fact,
it is a straightforward matter to verify that there are exactly~$8$ invariant subspaces~$n(\bN)$
with fixed spin content compatible with the motif~$(0,1,0,0)$ (i.e., with nonzero degeneracy),
whose respective degeneracies and energies are listed in Table~\ref{table.subs}.
\begin{table}[h]
  \centering
  \begin{tabular}{|c|c|c|}\hline
    $\bN$& Degeneracy& Energy\\
    \hline
    (1,2,2)& 1& $6J-2(\mu_1+\mu_2)$\\
    (3,0,2)& 1& $6J-2\mu_2$\\
    (3,2,0)& 1& $6J-2\mu_1$\\
    (4,0,1)& 1& $6J-\mu_2$\\
    (4,1,0)& 1& $6J-\mu_1$\\
    (2,1,2)& 2& $6J-(\mu_1+2\mu_2)$\\
    (2,2,1)& 2& $6J-(2\mu_1+\mu_2)$\\
    (3,1,1)& 3& $6J-(\mu_1+\mu_2)$\\
    \hline
  \end{tabular}
  \caption{Invariant subspaces~$n(\bN)$ compatible with the motif~$(0,1,0,0)$ in the $\su(1|2)$
    case.}
  \label{table.subs}
\end{table}

The above general results considerably simplify for the supersymmetric $\su(m)$ KY model without
magnetic field or chemical potential terms, given by Eq.~\eqref{tJSUSY}-\eqref{tKY}. Indeed, in
this case $h_\si=\mu_c=0$, so that from Eq.~\eqref{musim} we obtain
\[
  \mu_{\si}=t_0=\frac{t\pi^2}{3N^2}(N^2-1),\qquad 1\le\si\le m\,.
\]
Thus Eq.~\eqref{EdeN} becomes
\begin{equation}\label{EKY}
  \fl
  E_{\bde,\bN}=J\sum_{i=1}^{N-1}\de_i\cdot i(N-i)-t_0(N-N_0)
  =J\left[\sum_{i=1}^{N-1}\de_i\cdot i(N-i)-\frac16\,(N^2-1)(N-N_0)\right],
\end{equation}
which depends on~the spin content~$\bN$ only through~$N_0$. Thus for a given motif~$\bde$ all of
its compatible invariant subspaces~$n(\bN)$ with the same number of holes~$N_0$ will now have the
same energy. For instance, for the case~$N=5$, $m=2$ and the motif~$(0,1,0,0)$ considered above
there are only four sectors with different energies, corresponding to $N_0=1$ (singlet), $2$ (four
times degenerate), $3$ (five times degenerate) and~$4$ (twice degenerate), with respective
energies~$-2J(7-2N_0)$. In general, in order to compute the degeneracy associated to each
motif~$\bde$ in a sector with a given number of holes~$N_0$ we just have to count the number of
allowed Young tableaux according to rules i) and~ii) above, replacing rule three by the following:

\begin{description}
\item{iii')\en} The number~$0$ must appear exactly~$N_0$ times.
\end{description}
The rules i), ii), iii') provide a complete description of the spectrum of the
supersymmetric~$\su(m)$ KY model~\eqref{tJSUSY}-\eqref{tKY}, with the correct degeneracy for each
energy~\eqref{EKY}, for arbitrary values of $m$ and the number of sites~$N$. In particular, when
applied to the simplest $\su(2)$ case these rules provide the first rigorous proof known to the
authors to the long-standing conjecture of Saiga and Kuramoto mentioned in the Introduction.
\begin{table}[t] \centering
  \begin{tabular}{|c|c|c|c|c|c|c|c|}\hline
    Motif &0 &1 &2 &3 &4 &5 & Energy\\
    \hline
    $(0,0,0,0)$ &0 &0 &0 &1 &2 &1 & $-4J(5-N_0)$\\
    $(0,0,0,1$), $(1,0,0,0)$ &0 &0 &2 &4 &2 &0& $-4J(4-N_0)$\\
    $(0,0,1,0)$, $(0,1,0,0)$ &0 &1 &4 &5 &2 &0 & $-2J(7-2N_0)$\\
    $(1,0,0,1)$ &0 &4 &8 &4 &0 &0 &$-4J(3-N_0)$\\
    $(0,0,1,1)$, $(1,1,0,0)$ &0 &3 &6 &3 &0 &0 & $-2J(5-2N_0)$\\
    $(1,0,1,0)$, $(0,1,0,1)$ &2 &8 &10 &4 &0 &0 & $-2J(5-2N_0)$\\
    $(0,1,1,0)$ &2 &7 &8 &3 &0 &0 & $-4J(2-N_0)$\\
    $(1,0,1,1)$, $(1,1,0,1)$ &6 &12 &6 &0 &0 &0 & $-2J(3-2N_0)$\\
    $(0,1,1,1)$, $(1,1,1,0)$ &4 &8 &4 &0 &0 &0 & $-4J(1-N_0)$\\
    $(1,1,1,1)$ &6 &5 &0 &0 &0 &0 & $4JN_0$\\                                          
    \hline
  \end{tabular}
  \caption{Spectrum of the $\su(1|2)$ KY model~\eqref{tJSUSY}-\eqref{tKY} with~$N=5$ sites. The
    integers appearing in the columns labeled $0, \dots, 5$ are the degeneracies of the motif(s)
    in each row for the subspaces with $N_0=0, \dots, 5$ holes.}
  \label{table.spec}
\end{table}

As a simple illustration of the above assertion, we present in Table~\ref{table.spec} the detailed
spectrum of the (original) $\su(2)$ KY model~\eqref{tJSUSY}-\eqref{tKY} with~$N=5$ sites. More
precisely, in each of the columns of this table labeled with the integers $N_0=0,\dots,5$ we list
the degeneracy associated to the motif(s) in each row for the subspace with $N_0$ holes. This
degeneracy is computed by first generating all the skew Young tableaux compatible with each of the
$2^{N-1}=16$ possible motifs according to rules i), ii) and iii') above, which can be easily
accomplished using a simple \emph{Mathematica} program. We then sort the resulting tableaux
according to the number of holes (zeros) in each of them. By Eq.~\eqref{EKY}, a
motif~$\bde=(\de_1,\dots,\de_{N-1})$ and its reverse~$\bde'=(\de_{N-1},\dots,\de_1)$ clearly have
the same energy. In fact, it can be shown that two such motifs give rise to the same numbers of
compatible Young tableaux in each subspace~$n(\bN)$ (see the appendix for details). For this
reason, we have grouped together in Table~\ref{table.spec} two motifs that are the reverse of each
other. More generally, if we exchange any two components~$\de_k$ and~$\de_{N-k}$ of a motif~$\bde$
we obtain a motif with the same energy as~$\bde$ in each subspace with~$N_0$ holes. However, these
two motifs may not necessarily yield the same number of compatible Young tableaux in such a
subspace. For instance, the motifs $(0,0,1,1)$ and $(1,0,1,0)$ have the same energy~$-2J(5-2N_0)$
in a subspace with~$N_0$ holes, but it is clear from Table~\ref{table.spec} that their
degeneracies differ in each of these subspaces for $N_0=0,\dots,3$. Finally, it is apparent from
Table~\ref{table.spec} that the ground state is obtained from the motifs~$(0,0,1,0)$, $(0,1,0,0)$
(in the sector with one hole) and $(1,0,1,0)$, $(0,1,0,1)$ (in the sector with no holes). It is
thus six times degenerate, with energy $-10J$.

\section{Ground state phases for the supersymmetric spin~\mbox{\boldmath$1/2$} KY model}
\label{sec.GS}

The complete description of the spectrum of the supersymmetric KY model in terms of motifs, bond
vectors and their associated skew Young tableaux developed in the previous section is particularly
suited to studying its ground state. As an example, we shall compute next the ground state energy
per site of the~spin~$1/2$ KY model in the thermodynamic limit for all possible values of the
magnetic field strength~$h\equiv h_1$ and chemical potential~$\mu\equiv\mu_c+t_0$. To this end,
let us first choose the unit of energy so that $t=1/(2\pi^2)$, so that Eq.~\eqref{EdeN} reads
\begin{equation}\label{Ensu2}
E_{\bde,\bN}=\frac1{N^2}\sum_{i=1}^{N-1}\de_i\cdot i(N-i)-\frac h2\,(N_1-N_2)-\mu(N_1+N_2)\,.
\end{equation}
We shall also assume without loss of generality that~$h\ge0$, since taking $h<0$ simply reverses
the role of the ``up'' ($\si=1$) and ``down'' ($\si=2$) fermions. It is then clear from the term
proportional to~$h$ in Eq.~\eqref{Ensu2} that the ground state(s) must belong to an invariant
subspace~$n(\bN)$ with~$N_1\ge N_2$. Since the dispersion function~$i(N-i)$ is symmetric
about~$i=N/2$ and has an absolute maximum at this point, by Eq.~\eqref{specH} for $N$ large enough
the bond vector minimizing the energy in the subspace~$n(\bN)$ must be of the form\footnote{We are
  actually assuming here that both $N_1$ and~$N_2$ are even. In other cases the form of the
  minimizing bond vector differs slightly from Eq.~\eqref{de0}, but the formula for its energy
  coincides in the thermodynamic limit with the one given below. Note also that unless $N_1$
  and~$N_2$ are both odd there is actually an additional bond vector with the same energy
  as~\eqref{de0} (or its variants, for~$N_1$ and $N_2$ of opposite parity).}
\begin{equation}\label{de0}
  \bbs_0=(\,\underbrace{1\cdots1}_{\mathclap{(N_1-N_2)/2}}\,\underbrace{12\cdots12}_{N_2}\,
  \underbrace{0\cdots0}_{N_0}\,\underbrace{12\cdots12}_{N_2}\,
  \underbrace{1\cdots1}_{\mathclap{(N_1-N_2)/2}}\,)\,.
\end{equation}
By Eq.~\eqref{Ensu2}, the energy per site of the corresponding motif
\[
   \bde_0=(\,\underbrace{1,\dots,1,}_{\mathclap{(N_1-N_2)/2}}\,\underbrace{0,1,\dots,0,1,}_{N_2}\,
   \underbrace{0,\dots,0,}_{N_0}\,\underbrace{0,1,\dots,0,1,}_{N_2}\,
   \underbrace{1,\dots,1}_{\mathclap{(N_1-N_2)/2-1}}\,)
\]
is given by
\[
  \frac{E_{\bde_0,\bN}}N=\frac2N\sum_{k=1}^{Nt}\vep(x_k)
  +\frac2N\sum_{k=1}^{N(s-t)/2}\kern-6pt\vep(x_{Nt+2k})
  -ht-2\mu s\equiv u_{\bde_0,\bN}\,,
\]
where the factor of $2$ before the sums comes from the obvious symmetry of~$\bde_0$ around $N/2$
and we have set
\[
  s=\frac{N_1+N_2}{2N},\qquad t=\frac{N_1-N_2}{2N},\qquad x_k=\frac{k}{N}\,,
  \qquad \vep(x)=x(1-x)\,.
\]
Hence in the thermodynamic limit the minimum energy per site in a subspace~$n(\bN)$ with
$N_1\ge N_2$ is given by
\begin{eqnarray}
  \fl
  u(s,t)\equiv\lim_{N\to\infty}u_{\bde_0,\bN}
  &=2\int_0^t\vep(x)\,\diff x+\int_t^s\vep(x)\,\diff
    x-ht-2\mu s\nonumber\\
  \fl
  &=\int_0^s\vep(x)\,\diff x+\int_0^t\vep(x)\,\diff x-ht-2\mu s
    \equiv f_{2\mu}(s)+f_h(t)\,,
    \label{ude0}
\end{eqnarray}
where
\[
  f_\la(s)=\int_0^s\vep(x)\,\diff x-\la s\equiv f(s)-\la s\,.
\]
The ground state energy of the spin~$1/2$ KY model in the thermodynamic limit is thus the minimum
value of the function $u(s,t)$ in the triangle
\[
  D=\{(s,t)\in\RR^2:0\le t\le s\le 1/2\}\,.
\]
In order to compute this minimum value, note first of all that~$f'_\la(x)=\vep(x)-\la$, with
$\vep(x)$ monotonically increasing from~$0$ to $1/4$ in the interval~$[0,1/2]$. Hence $f_\la$ is
monotonically increasing\footnote{For the sake of conciseness, we shall implicitly assume in what
  follows that the functions~$f_\la$ and~$\vep$ are restricted to the interval of
  interest~$[0,1/2]$.} over the interval $[0,1/2]$ for~$\la\le0$, monotonically decreasing
for~$\la\ge1/4$, and has a unique global minimum at the point
\[
  x_0(\la)\equiv\vep^{-1}(\la)=\frac12\,\big(1-\sqrt{1-4\la}\,\big)\in(0,1/2)
\]
for $0<\la<1/4$. We thus have the following possibilities (recall that we are assuming throughout
that $h\ge0$):

\medskip

\noindent i)\en $\dfrac h2<\mu<1/8$

\smallskip
\noindent In this case~$f_h(t)$ and $f_{2\mu}(s)$ both have a unique minimum over the
interval~$[0,1/2]$ respectively at the points~$t_0=x_0(h)\in[0,1/2)$
and~$s_0=x_0(2\mu)\in(0,1/2)$, with~$t_0< s_0$ since~$\vep(t_0)=h<2\mu=\vep(s_0)$. Hence~$u$
attains its global minimum on~$D$ at the point~$(s_0,t_0)$ (which is an interior point if~$h>0$),
and the ground state energy density~$u$ is consequently given by
\begin{eqnarray}
  u&=u(s_0,t_0)=\int_0^{x_0(h)}\vep(x)\,\diff x+\int_0^{x_0(2\mu)}\vep(x)\,\diff x-hx_0(h)-2\mu
     x_0(2\mu)\nonumber\\
   &=\frac1{6}\,\bigg[1-3(h+2\mu)-\frac12(1-4h)^{3/2}-\frac12(1-8\mu)^{3/2}\bigg]\,.
     \label{uT}
\end{eqnarray}
Note also that $t=(N_1-N_2)/(2N)$ and~$2s=(N_1+N_2)/N$ are respectively equal to the magnetization
and the charge density per site (assuming that the fermions have unit charge and gyromagnetic
ratio equal to~2). Hence the zero-temperature magnetization and charge densities are given by
\[
  m_s=x_0(h)=\frac12\,\big(1-\sqrt{1-4h}\,\big)\,,\qquad
  n_c=2x_0(2\mu)=1-\sqrt{1-8\mu}\,.
\]
The corresponding magnetic and charge susceptibilities are obtained by differentiation, namely,
\[
  \fl
  \chi_s=\pdf{m_s}h=(1-4h)^{-1/2}=(1-2m_s)^{-1}\,,\quad
  \chi_c=\pdf{n_c}\mu=4(1-8\mu)^{-1/2}=4(1-n_c)^{-1}\,,
\]
in agreement with known results (see, e.g.,~\cite{KK95}).

\medskip\noindent

If $(h,\mu)$ lies outside the region~$\{(h,\mu)\in\RR^2:0\le h/2<\mu<1/8\}$, the system
\[
  \vep(s)=2\mu\,,\qquad \vep(t)=h
\]
determining the critical points of $u$ has no solutions within the interior of~$D$. Thus the
function~$u(s,t)$ must necessarily attain its minimum value in the triangle~$D$ on its sides. It
is worth noting in this respect that this minimum cannot be reached at an interior point of the
horizontal side~$t=0$ unless~$h=0$, since for~$h>0$ we have~$f_h'(0)=-h<0$ and thus
\[
  u(s,0)=f_{2\mu}(s)>f_{2\mu}(s)+f_h(t)=u(s,t)
\]
for sufficiently small~$t>0$. We are left with the following possibilities:

\medskip
\noindent ii)\en $h\ge1/4$, $\mu+\dfrac h2\ge1/4$

\smallskip\noindent In this case $u(s,s)=f_h(s)+f_{2\mu}(s)=2f_{h/2+\mu}(s)$
and~$u(1/2,t)=f_h(t)+f_{2\mu}(1/2)$ are both decreasing, so that these functions have a unique
global minimum on the interval~$[0,1/2]$ at the right endpoint~$1/2$, with common value
\begin{equation}\label{uminW1}
  u(1/2,1/2)=2f_{h/2+\mu}(1/2)=
  \frac1{6}\,\big[1-3(h+2\mu)\big]\,.
\end{equation}
This must be the unique minimum of~$u$ on~$D$, since as~$h>0$ this minimum cannot be attained on
the interior of the side~$t=0$. Hence in this case the ground state energy per site is given by
Eq.~\eqref{uminW1}, while the zero-temperature magnetization and charge densities are simply
\[
  n_c=2m_s=1\,.
\]
Note that~$s=t=1/2$ is equivalent to~$N_1=N$, so that this phase consists only of up fermions.

\medskip
\noindent iii)\en $\mu+\dfrac h2\le0$

\smallskip\noindent Since~$\mu\le -h/2\le0$, the minimum of~$u(s,0)=f_{2\mu}(s)$
and~$u(s,s)=2f_{h/2+\mu}(s)$ on~$[0,1/2]$ is located at~$s=0$, while
\[
  u(1/2,t)=f(t)+f(1/2)-ht-\mu> -\frac h2-\mu\ge 0=u(0,0)\,.
\]
Thus in this case the ground state energy, zero-temperature magnetization and charge all vanish.
In fact, since $s=t=0$ is equivalent to $N_0=N$, this is the trivial phase consisting only of
holes.

\medskip
\noindent iv)\en $h<1/4$, $\mu\ge1/8$

\smallskip\noindent In this case $u(1/2,t)$ attains its minimum value in the interval~$[0,1/2]$ at
$t_0=x_0(h)\in[0,1/2)$. On the other hand,
\[
  u(s,s)=f_h(s)+f_{2\mu}(s)> f_h(t_0)+f_{2\mu}(1/2)=u(1/2,t_0),
\]
since $f_{2\mu}$ is decreasing in the interval~$[0,1/2]$ on account of the condition~$\mu\ge1/8$.
Similarly,
\[
  u(s,0)=f_{2\mu}(s)\ge f_{2\mu}(1/2)=u(1/2,0)\ge u(1/2,t_0),
\]
with equality only if~$h=0$ and~$s=1/2$. Thus the global minimum of~$u$ on~$D$ is attained at the
point~$(1/2,t_0)$. In particular, the ground state energy per site is given by
\begin{eqnarray}
  u=u(1/2,t_0)&=\int_0^{x_0(h)}\vep(x)\,\diff x-hx_0(h)+\int_0^{1/2}\vep(x)\,\diff x-\mu\nonumber\\
  &= \frac1{6}\,\bigg[1-3(h+2\mu)-\frac12(1-4h)^{3/2}\bigg]\,,
      \label{uminB1}
\end{eqnarray}
while the magnetization and charge per site at zero temperature read
\[
  m_s=x_0(h)\,,\qquad n_c=1\,.
\]
The $T=0$ magnetic susceptibility is again
\[
  \chi_s=(1-4h)^{-1/2}=(1-2m_s)^{-1}\,,
\]
while the charge susceptibility vanishes. Since~$s=1/2$ is equivalent to~$N_1+N_2=N$, this is
an~$\su(2)$ phase characterized by the absence of holes.

\medskip
\noindent v)\en $0< h/2+\mu<1/4$, $\mu\le h/2$

\smallskip\noindent The above inequalities imply that $u(s,s)=2f_{h/2+\mu}(s)$ attains its unique
global minimum over the interval~$[0,1/2]$ at the point~$s_0=x_0(h/2+\mu)\in(0,1/2)$. It is
straightforward to check that~$u(s,t)$ achieves its minimum over the domain~$D$ at the
point~$(s_0,s_0)$. Indeed, as $h>0$ in this case this minimum cannot be reached on the interior of
the side~$t=0$. Consider next the side~$s=1/2$. If~$h\ge1/4$ the function~$f_h(t)$ is decreasing,
so that
\[
  \fl
  u(1/2,t)=f_h(t)+f_{2\mu}(1/2)\ge f_h(1/2)+f_{2\mu}(1/2)=u(1/2,1/2)>u(s_0,s_0).
\]
On the other hand, if $0\le h<1/4$ then~$f_h$ has a global minimum on~$[0,1/2]$
at~$t_0=x_0(h)\ge x_0(h/2+\mu)=s_0$, with~$t_0\in[0,1/2)$. Since $\vep(x)-2\mu>0$ for~$x> t_0$
(this is obvious for~$\mu\le0$, while for~$0<\mu\le1/8$ it is a consequence of the
inequality~$x_0(2\mu)\le x_0(h)$, which in turn follows from $2\mu\le h$) we have
\[
  f_{2\mu}(1/2)=\int_{0}^{\mathrlap{1/2}}\big(\vep(x)-2\mu\big)\,\diff x>
  \int_{0}^{\mathrlap{t_0}}\big(\vep(x)-2\mu\big)\,\diff x=f_{2\mu}(t_0)\,.
\]
\begin{table}[t]
  \centering\small
  \begin{tabular}{|c|c|c|}\hline
    Region& 
            Ground state energy per site& Spin content\\ \hline\hline
       $\frac h2<\mu<\frac 18$& $\frac1{6}\big[1-3(h+2\mu)-\frac12(1-4h)^{3/2}
                                -\frac12(1-8\mu)^{3/2}\big]$&$B,F_1,F_2$\vru\\ \hline
    $h\ge\frac 14$,\quad $\mu+\frac h2\ge\frac 14$
          & $\frac1{6}\,\big[1-3(h+2\mu)\big]$&$F_1$\vru\\ \hline
        $\mu+\frac h2\le0$& $0$&$B$\vru\\ \hline
        $h< \frac 14$,\quad $\mu\ge\frac 18$&$\frac1{6}\big[1-3(h+2\mu)-\frac12(1-4h)^{3/2}\big]$
                                          &$F_1,F_2$\vru\\ \hline
$\,0<\mu+\frac h2<\frac 14$,\quad $\mu\le\frac h2\,$
          & $\frac16\Big[1-3(h+2\mu)-(1-2h-4\mu)^{3/2}\Big]$&$B,F_1$\vru\\ \hline
  \end{tabular}
  \caption{Ground state energy per site and spin content ($B\equiv$ hole (boson),
    $F_\si\equiv$ fermion of type~$\si$) for the spin~$1/2$ KY model as a function of its
    parameters~$h\ge0$ and $\mu$ (the unit of energy has been taken as
    $2\pi^2t$).}\label{table.regs}
\end{table}
Hence
\begin{eqnarray*}
  \fl
  u(s_0,s_0)-u(1/2,t_0)&=2f_{h/2+\mu}(s_0)-f_h(t_0)-f_{2\mu}(1/2)\\&<
  2f_{h/2+\mu}(s_0)-f_h(t_0)-f_{2\mu}(t_0)=2f_{h/2+\mu}(s_0)-2f_{h/2+\mu}(t_0)\le0\,,
\end{eqnarray*}
which completes the proof of our assertion. In summary, in this case the ground state energy per
site is given by
\begin{eqnarray}
  u=u(s_0,s_0)&=2\int_0^{\mathrlap{x_0(h/2+\mu)}}
  \vep(x)\,\diff x-(h+2\mu)x_0\bigl(\tfrac h2+\mu\bigr)\nonumber\\
  &=\frac16\Big[1-3(h+2\mu)-(1-2h-4\mu)^{3/2}\Big],
  \label{uminB0}
\end{eqnarray}
while the magnetization, the charge density and their susceptibilities (per site) read
\begin{eqnarray*}
  &n_c=2m_s=2x_0(h/2+\mu)=1-\sqrt{1-2h-4\mu}\,,\\
  &\chi_c=4\chi_s=2(1-2h-4\mu)^{-1/2}=\frac2{1-n_c}=\frac{2}{1-2m_s}\,.
\end{eqnarray*}
Note, finally, that the equality~$s=t$ is equivalent to~$N_2=0$. This is thus an~$\su(1|1)$ phase,
consisting only of holes and up fermions.

\begin{figure}[t]
  \centering
  \includegraphics[width=.46\textwidth]{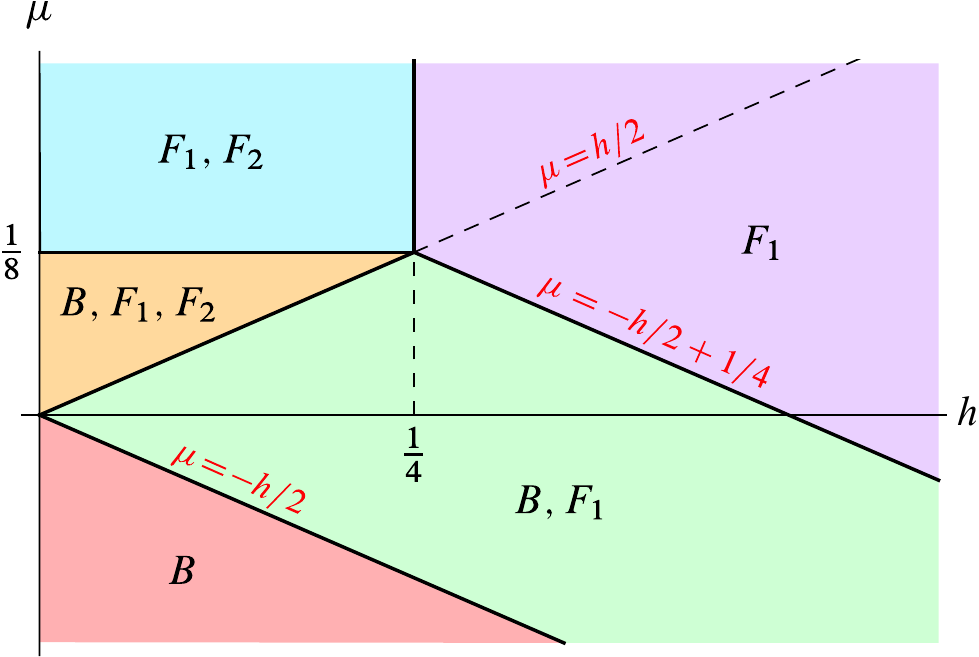}\hfill
  \includegraphics[width=.46\textwidth]{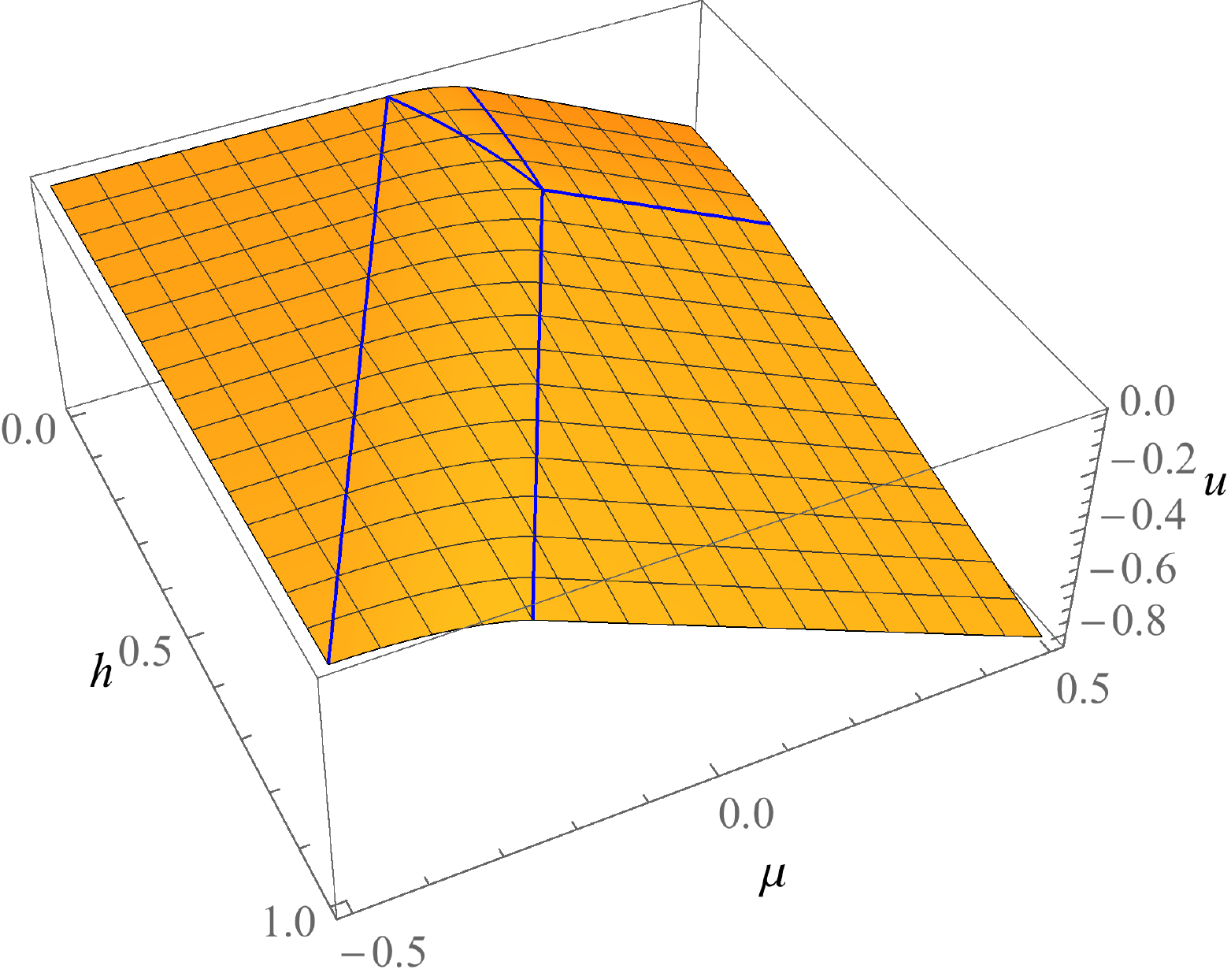}\hfill
  \caption{Left: phase diagram of the ground state of the spin~$1/2$ KY model ($B\equiv$ hole
    (boson), $F_\si\equiv$ fermion of type~$\si$). Right: ground state energy per site of the
    latter model as a function of its parameters~$h\ge0$ and $\mu$ (the blue lines represent the
    boundaries of the regions in the phase diagram). In both plots, the unit of energy has been
    taken as~$2\pi^2t$.}
  \label{fig.u}
\end{figure}

Our results are summarized in Table~\ref{table.regs}. Note, in particular, that $u$ is continuous
(indeed, of class $C^1$) over its domain, although its second derivatives are discontinuous along
the boundaries of the regions listed in Table~\ref{table.regs} (cf.~Fig.~\ref{fig.u}).

\section{Conclusions}\label{sec.conc}

Although the supersymmetric $\su(m)$ $t$-$J$ model with long-range interactions, also known as the
$\su(m)$ KY model, has been studied extensively during the last few decades, an analytical
derivation of its spectrum and exact partition function has been missing so far. With the purpose
of filling up this gap, in this paper we first establish the precise equivalence of the $\su(m)$
KY model to a suitable modification of the $\su(1|m)$ HS spin chain with chemical potential terms.
This equivalence allows us to obtain the partition function of the former model from that of the
latter, which was recently computed in Ref.~\cite{FGLR18}. A remarkable property of this partition
function is the fact that it can be rewritten as the partition function of a suitable
inhomogeneous vertex model. 
Analyzing the structure of these two equivalent partition functions, we not only obtain the
complete spectrum of the $\su(m)$ KY model in the presence of an arbitrary magnetic field and
charge chemical potential, but also develop a novel method for determining the magnon numbers or
spin content of the corresponding wave functions. This yields an exhaustive description of the
spectrum in the subspaces with well-defined magnon content in terms of suitably restricted bond
configurations of the equivalent vertex model, which are closely connected with supersymmetric
versions of Haldane's motifs and their related skew Young tableaux. For the particular case $m=2$,
this description provides a rigorous proof of a long-standing conjecture by Saiga and
Kuramoto~\cite{SK99} based on numerical evidence.

As a concrete application of our results, we study various thermodynamic properties of the
$\su(2)$ KY model in the zero temperature limit. To this end, we determine the structure of the
motifs and bond configurations yielding the ground state of the latter model in the thermodynamic
limit for different values of the external parameters. The structure of such bond configurations
leads to a complete description of the different ground state phases in terms of the magnetic
field strength and the charge chemical potential. These phases are characterized by the spin
content (magnon numbers) of the corresponding wave functions, namely an $\su(1|2)$ phase where
holes and fermions of both types co-exist, an $\su(1|1)$ phase with holes and fermions of only one
type, and an $\su(0|2)$ phase with fermions of both types, apart from the trivial phases
consisting of only holes or fermions of one type. We also compute the zero-temperature values of
the energy, magnetization and charge density, along with the magnetic and charge susceptibilities,
for each ground state phase. This description of the thermodynamic properties at zero temperature
goes beyond previously known results, which were derived by different methods and restricted to
the $\su(1|2)$ phase. In particular, our analysis confirms that the strong spin-charge separation
characteristic of the long-range $t$-$J$ model at low temperatures~\cite{KK95} occurs in all
nontrivial phases.

Note, finally, that the description of the spectrum of the~$\su(m)$ KY model in terms of
supersymmetric motifs and their associated Young tableaux derived in this paper makes it possible
to compute in closed form the model's thermodynamic functions at finite temperature, by means of
the transfer matrix method developed in Refs.~\cite{EFG12,FGLR18}. In fact, work on this problem
is currently in progress and shall be presented in a forthcoming publication.

\section*{Acknowledgments}
We would like to thank the anonymous referee, whose remarks helped to improve the presentation.
This work was partially supported by Spain's MINECO grant~FIS2015-63966-P. JAC would also like to
acknowledge the financial support of the Universidad Complutense de Madrid through a 2015
predoctoral scholarship.

\appendix

\section{Degeneracy of a reverse motif}
\label{app.KK96}

In this appendix we shall show that the degeneracy~$d(\bde,\bN)$ of the
motif~$\bde=(\de_1,\dots,\de_{N-1})$ in the subspace~$n(\bN)$, i.e., the number of allowed Young
tableaux for this motif containing $N_\al$ instances of each of the integer~$\al\in\{0,\dots,m\}$,
is equal to that of its reverse~$\bde'\equiv(\de_{N-1},\dots,\de_1)$. To this end, define the
super-Schur polynomial associated to the motif~$\bde$ by
\[
  S_{\bde}(x,\by)=\sum_{T\in\bde} x^{N_0(T)}y_1^{N_1(T)}\cdots\, y_m^{N_m(T)}\,,
\]
where the sum is extended to all allowed Young tableaux~$T$ associated to~$\bde$ according to
rules~i)--iii) in Section~\ref{sec.motifs}, and~$N_\al(T)$ denotes the number of times the
integer~$\al$ appears in~$T$. It is well known (see, e.g.,\cite{KKN97,HB00,BBHS07}) that this
polynomial can be computed from the determinantal formula
\[
  S_{\bde}=
  \left|
  \begin{array}{ccccc}
    E_{k_r}& E_{k_{r-1}+k_r}& E_{k_{r-2}+k_{r-1}+k_r}&\cdots &E_{k_1+\cdots+k_r}\\
    1& E_{k_{r-1}}& E_{k_{r-2}+k_{r-1}}&\cdots &E_{k_1+\cdots+k_{r-1}}\\
    0& 1& E_{k_{r-2}}& \cdots& E_{k_1+\cdots+k_{r-2}}\\
    \vdots&\vdots& \vdots& \vdots&\vdots\\
   0& \cdots& 1& E_{k_2}& E_{k_1+k_2}\\
    0&\cdots& 0& 1& E_{k_1}
  \end{array}
  \right|\equiv S\langle k_1,\dots,k_r\rangle\,,
\]
where $\sum_{j=1}^ik_j$ (with $1\le i\le r-1$) denotes the position of the~$i$-th $1$ in the
motif~$\bde$,
\[
  k_r=N-\sum_{i=1}^{r-1}k_i\,,
\]
and the polynomials~$E_k(x,\by)$ are defined in terms of the elementary symmetric
polynomials~\eqref{ejdef} by
\[
  E_k(x,\by)=\sum_{l=0}^k x^{k-l}e_l(\by)\,.
\]
Clearly, for the reverse motif~$\bde'$ we have~$k'_i=k_{r+1-i}$, so that
\[
  S_{\bde'}=S\langle k_r,\dots,k_1\rangle\,.
\]
Since, by definition of $S_{\bde}$, $d(\bde,\bN)$ is the coefficient of
$x^{N_0}y_1^{N_1}\cdots y_m^{N_m}$ in $S_{\bde}$, to prove the equality of $d(\bde,\bN)$ and
$d(\bde',\bN)$ it suffices to show that
\begin{equation}\label{Skkp}
  S\langle k_1,\dots,k_r\rangle=S\langle k_r,\dots,k_1\rangle\,.
\end{equation}
We shall establish the latter equality by induction on~$r$. To begin with, note that~\eqref{Skkp}
is trivially obvious for~$r=1$. Suppose now that the latter equation holds for
determinants~$S\langle\cdot\rangle$ of order up to $r-1$.
Expanding~$S\langle k_1,\dots,k_r\rangle$ by the first column we obtain:
\begin{equation}\label{Sexpcol}
  S\langle k_1,\dots,k_r\rangle=E_{k_r}S\langle k_1,\dots,k_{r-1}\rangle-
  S\langle k_1,\dots,k_{r-2},k_{r-1}+k_r\rangle\,.
\end{equation}
Similarly, expanding $S\langle k_r,\dots,k_1\rangle$ by the last row we have
\begin{equation}\label{Sexprow}
  S\langle k_r,\dots,k_1\rangle=E_{k_r}S\langle k_{r-1},\dots,k_{1}\rangle-
  S\langle k_{r-1}+k_r,k_{r-2},\dots,k_{1}\rangle\,.
\end{equation}
Equation~\eqref{Skkp} follows immediately from Eq.~\eqref{Sexpcol} using the induction hypothesis
and Eq.~\eqref{Sexprow}. Note, finally, that the latter proof can be easily adapted to the
$\su(n|m)$ case with arbitrary $n$.

\section*{References}


\end{document}